\documentclass[12pt]{article}

\usepackage{graphicx}
\usepackage[body={17.5cm, 21cm},right=2cm]{geometry}
\usepackage{amssymb}
\usepackage{amsmath}

\newcommand\ee{\end{equation}}
\newcommand\be{\begin{equation}}
\newcommand\eea{\end{eqnarray}}
\newcommand\bea{\begin{eqnarray}}
\newcommand{\sfrac}[2]{{\textstyle\frac{#1}{#2}}}
\newcommand\di{\partial}
\newcommand\mpl{M_{\rm Pl}}
\def\la{\langle}
\def\ra{\rangle}
\def\beq{\begin{equation}}
\def\eeq{\end{equation}}
\def\d{\partial}

\begin{document}
\setcounter{page}{0}
\thispagestyle{empty}

\begin{titlepage}

\begin{center}

~

\vspace{1.cm}

{\LARGE \sc{
A Measure of de~Sitter Entropy \\[.5cm] and Eternal Inflation
}}\\[1cm]
{\large Nima Arkani-Hamed$^{\rm a}$, Sergei Dubovsky$^{\rm a,b}$, Alberto Nicolis$^{\rm a}$, \\[.2cm] Enrico Trincherini$^{\rm a}$,
and Giovanni Villadoro$^{\rm
a}$}
\\[0.6cm]

{\small \textit{$^{\rm a}$ Jefferson Physical Laboratory, \\ Harvard University, Cambridge, MA 02138, USA}}

\vspace{.2cm}
{\small \textit{$^{\rm b}$ Institute for Nuclear Research of the Russian Academy of Sciences, \\
        60th October Anniversary Prospect, 7a, 117312 Moscow, Russia}}

\end{center}

\vspace{.8cm}
\begin{abstract}
We show that in any model of non-eternal inflation satisfying the null energy condition, the area of the de~Sitter horizon
increases by at least one Planck unit in each inflationary $e$-folding. This observation gives an
operational meaning to the finiteness of the entropy $S_{\rm dS}$ of an inflationary de~Sitter space eventually exiting into an asymptotically flat region: the
asymptotic observer is never able to measure more than $e^{S_{\rm dS}}$ independent
inflationary modes. This suggests a limitation  on the amount of de~Sitter space outside the horizon
that can be consistently described at the semiclassical level,  fitting well with other examples of the breakdown of locality in quantum gravity, such as in black hole evaporation.
The bound does not hold in models of inflation that violate the null energy condition, such as ghost inflation. This strengthens the  
case for the thermodynamical interpretation of the bound as conventional black hole thermodynamics also fails in these models, strongly suggesting that 
these theories are incompatible with basic gravitational principles.  

\end{abstract}

\end{titlepage}

\section{Introduction}
String theory appears to have a landscape of vacua \cite{Bousso:2000xa,Douglas:2006es}, and eternal inflation \cite{Vilenkin:1983xq,Linde:1986fd}
is a plausible mechanism for populating them. In this picture there is an 
infinite volume of spacetime undergoing eternal inflation, nucleating 
bubbles of other vacua that either themselves eternally inflate, or end in 
asymptotically flat or AdS crunch space-times. These different regions 
are all space-like separated from each other and are therefore naively completely 
independent. The infinite volumes and infinite numbers of bubbles vex 
simple attempts to define a ``measure" on the space of vacua, since these 
involve ratios of infinite quantities.

This picture relies on an application of low-energy effective field 
theory to inflation and bubble nucleation. On the face of it this is 
totally justified, since everywhere curvatures are low compared to the 
Planck or string scales. However, we have long known that effective field 
theory can break down dramatically even in regions of low curvature, indeed 
it is precisely the application of effective field theory within its 
putative domain of validity that leads to the black hole information 
paradox. Complementarity \cite{'tHooft:1990fr,Susskind:1993if} suggests that regions of low-curvature spacetime 
that are space-like separated may nonetheless not be independent. How can 
we transfer these relatively well-established lessons to de Sitter space and eternal inflation \cite{Banks:2001yp}? 

In this note we begin with a brief discussion of why locality is 
necessarily an approximate concept in quantum gravity, and why the failure 
of locality can sometimes manifest itself macroscopically as in the 
information paradox (see, e.g.,
\cite{Banks:2002wr, Giddings:2005id, Giddings:2007ie} for  related discussions with somewhat different accents). Much of this material is review, though some of the emphasis is novel. The conclusion is simple: effective field theory breaks down when it 
relies on the presence of $e^{S}$ states behind a horizon of entropy $S$. 
Note that if the spacetime geometry is kept fixed as gravity is decoupled 
$G \to 0$, the entropy goes to infinity and effective field theory is a 
perfectly valid description.
In attempting to extend these ideas to de~Sitter space, there is a basic 
confusion. It is very natural to assign a finite number of states to a 
black hole, since it occupies a finite region of space \cite{Bekbound}. De~Sitter space also has a finite entropy \cite{Gibbons:1977mu}, but 
its spatially flat space-like surfaces have infinite volume, and it is not completely clear what this 
finite entropy means operationally, though clearly it must be associated with the fact that any given observer only sees a finite volume of de Sitter space. We regulate this question by considering approximate 
de~Sitter spaces which are non-eternal inflation models, exiting into 
asymptotically flat space-times. We show that for a very broad class of 
inflationary models, as long the null-energy condition is satisfied, the 
area of the de~Sitter horizon grows by at least one Planck unit during each 
$e$-folding, so that $dS_{\rm dS}/dN_e \gg 1$, and so the number of $e$-foldings of 
inflation down to a given value of inflationary Hubble is bounded as $N_e 
\ll S_{\rm dS}$ (limits on the effective theory of inflation have also been considered in e.g.~\cite{Albrecht:2002xs, Banks:2003pt}). This provides an operational meaning to the finiteness of the 
de~Sitter entropy: the asymptotic observer detects a spectrum of 
scale-invariant perturbations that she associates with the early de~Sitter 
epoch; however, she never measures more than $e^{S_{\rm dS}}$ of these modes. 
The bound is violated when the conditions for eternal inflation are 
met; indeed, $dS_{\rm dS}/dN_e \lesssim 1$ thereby provides a completely macroscopic 
characterization of eternal inflation. This bound suggests that no more 
than $e^{S_{\rm dS}}$ spacetime Hubble volumes can be consistently described 
within an effective field theory.

Our bound does not hold in models of inflation that violate the 
null-energy condition. Of course most theories that violate this energy 
condition are obviously pathological, with instabilities present even at 
the long distances. However in the last number of years, a 
class of theories have been studied \cite{ghost_C,Rubakov:2004eb,Dubovsky:2004sg}, loosely describing various ``Higgs" 
phases of gravity, which appear to be consistent as long-distance 
effective theories, and which (essentially as part of their {\it raison 
d'etre}) violate the null energy condition. Our result suggests that these 
theories violate the thermodynamic interpretation of de~Sitter entropy---an 
asymptotic observer exiting into flat space from ghost inflation \cite{ghost_I} could, for 
instance, measure parametrically more than $e^{S_{\rm dS}}$ inflationary modes. 
This fits nicely with other recent investigations \cite{Dubovsky:2006vk,Eling:2007qd}
that show that the second law 
of black hole thermodynamics also fails for these models. Taken together 
these results strongly suggest that, while these theories may be consistent 
as effective theories, they are in the ``swampland" of effective theories 
that are incompatible with basic gravitational principles \cite{Vafa:2005ui,Arkani-Hamed:2006dz}.


\section{Locality, gravity, and black holes}

\subsection{Locality in gravity}\label{sec:locality}

Since the very early days of quantum gravity it has been appreciated that the notion of local off-shell
observables is not sharply well-defined (see e.g. \cite{DeWitt:1967ub}).
It is important to realize that
it is {\it dynamical} gravity that is crucial for this conclusion, and not just the reparameterization invariance
of the theory. The existence of local operators
clashes with causality in a theory with a {\it dynamical} metric. Indeed, causality tells that the commutator of local operators
taken at space-like separated points should be zero,
\[
\left[ {\cal O}(x),{\cal O}(y)\right]=0 \quad \mbox{ if } \; (x-y)^2>0
\]
However, whether two points are space-like separated or not is determined by the metric, and is
not well defined if the metric itself fluctuates. Clearly, this argument crucially relies on the ability
of the metric to fluctuate, {\it i.e.} on the non-trivial dynamics of gravity.
Another argument is that Green's functions of local field operators, such as 
$\langle \phi(x) \phi(y)\cdots\rangle$ are not {\it invariant} (as opposed to {\it covariant}) under
coordinate changes. Consequently, they cannot represent physical quantities in a theory of gravity, where
coordinate changes are gauge transformations. Related to this, there is no standard notion of time evolution in gravity. Indeed, as a consequence of time reparameterization invariance,
the canonical quantization of general relativity leads to the Wheeler-de Witt equation~\cite{WdW},
which is analogous to the Schroedinger equation in ordinary quantum mechanics, but does not involve time,
\be
\label{WDW}
{\cal H}\Psi=0
\ee

These somewhat formal arguments seem to rely only on the reparametrization invariance of the theory, but of course this is incorrect---it is the dynamical gravity that is the culprit. To see this,  let us take the decoupling limit $M_{\rm Pl}\to
\infty$, so that gravity becomes non-dynamical. If we are in flat space, in this limit the metric $g_{\alpha \beta}$ must be diffeomorphic to $\eta_{\alpha \beta}$:
\be
\label{gold}
g_{\alpha\beta}=\frac{\d Y^\mu}{\d x^\alpha}\frac{\d Y^\nu}{\d x^\beta}\eta_{\mu\nu}
\ee
where $\eta_{\mu\nu}$ is the Minkowski metric and $Y^\mu$'s are to be thought of as the component functions of the space-time diffeomorphism (diff),
$x^\mu\to Y^\mu(x)$. The resulting theory is still reparameterization invariant, with matter fields
transforming in the usual way under the space-time diffs $x\to\xi(x)$, and the transformation rule
of the $Y^\mu$ fields is
\[
Y^\mu\to\left(\xi^{-1}\circ Y\right)^\mu
\]
where $\circ$ is the natural multiplication of two diffeomorphisms. Nevertheless, there are
local diff-invariant observables now, such as $\langle\phi(Y(x))\phi(Y(y))\cdots\rangle$.
Of course this theory is just equivalent to the conventional flat space field theory, which is recovered
in the ``unitary" gauge $Y^\mu=x^\mu$. Conversely, any field theory can be made diff invariant 
by introducing the ``Stueckelberg" fields $Y^\mu$ according to (\ref{gold}).
Diff invariance by itself, like any gauge symmetry, is just a redundancy of the description and
cannot imply any physical consequences. Conventional time evolution is also recovered in the decoupling limit; the Hamiltonian constraint (\ref{WDW}) still holds as a consequence of time reparameterization invariance, and in the decoupling limit the Hamiltonian ${\cal H}$ is 
\be
\label{Shroed}
\left[\frac{\d Y^\mu}{\d x^0}p_\mu+H_M\right]\Psi[{Y^\mu,{ matter}}]=0
\ee
where $H_M$ is the matter Hamiltonian.
Noting that the canonical conjugate momenta act as
\[
p_\mu=i\frac{\d}{\d Y^\mu}
\]
we find that the Hamiltonian constraint reduces to the conventional time-dependent
Schroedinger equation with $\Psi$ depending on time through $Y^\mu$. In a sense,
the gauge degrees of freedom $Y^\mu$'s play the role of clocks and rods in the decoupling limit.

This is to be contrasted with what happens for finite $M_{\rm Pl}$. In this case it is not possible to
explicitly disentangle the gauge degrees of freedom from the metric.
As a result to recover the conventional time evolution from the
Wheeler-de Witt equation one has to specify some physical clock field
(for instance, it can be the scale factor of the Universe, or some rolling scalar field), and use this field
similarly to how we used $Y^\mu$'s in (\ref{Shroed}) to recover the time-dependent Schroedinger
equation \cite{Lapchinsky:1979fd, Banks:1984cw, Banks:1984np}.
This strongly suggests that with dynamical gravity  one is forced to consider whether there exist physical
clocks that can resolve a given physical process.  In particular, this means
that in a region of size $L$ it does not make sense to discuss time evolution with
resolution better than $\delta t\sim (LM_{\rm Pl}^2)^{-1}$, as any physical clocks aiming to measure
time with that precision by the uncertainty principle would collapse the whole region into a black hole.

What does the formal absence of local observables in gravity mean operationally? There must be an intrinsic obstacle to measuring local observables with arbitrary precision; what is this intrinsic uncertainty? 
Imagine we want to determine the value of the 2-point function $\la\phi(x)\phi(y)\ra$
of a scalar field $\phi(x)$ between two space-like separated points $x$ and $y$.
We have to set up an apparatus that measures $\phi(x)$ and $\phi(y)$, repeat the
experiment $N$ times and collect the outcomes $\phi_i(x)$, $\phi_i(y)$ 
for $i = 1, \cdots, N$. We can then plot the values for the product 
$\phi_i(x) \phi_i(y)$, which will be peaked around some value. 
%
%
The width of the distribution will represent the uncertainty due to quantum fluctuations.
Without gravity there is no limit to the precision we can reach, just
by increasing $N$ the width of the distribution decreases as $1/\sqrt{N}$.
The presence of gravity, however, sets an intrinsic systematic uncertainty
in the measurement. The Bekenstein bound~\cite{Bekbound}, indeed, limits the number of 
states in a localized region of space-time.  This is due to the fact that,
in a theory with gravity, the object with the largest density 
of states is a black hole, whose size $R_S$ grows with its entropy ($S_{\rm BH}=R_S^{D-2}/4G$), 
or equivalently, with the number of states it can contain ($\sim e^{S_{\rm BH}}$). 
This means that an apparatus of finite size has a finite number 
of degrees of freedom (d.o.f.), thus can reach only a finite precision, 
limited by the number of states. 
For an apparatus with size smaller than $r=|x-y|$, the number of d.o.f. is bounded
by $S=r^{D-2}/G$. 
Without gravity there is no limit to the number of d.o.f. a compact apparatus can have so that the indetermination in the two-point function is only limited by the statistical error, which can be reduced indefinitely by increasing the number of measurements $N$. With gravity instead this is no longer true; an intrinsic systematic error (which must be a decreasing function of $S$) is always present to fuzz the notion of locality. The only two ways to eliminate such indetermination 
are: a)~by switching off gravity ($G\to0$);
b)~by giving up with local observables and considering only 
$S$-matrix elements (for asymptotically Minkowski spaces) where $r\to \infty$: 
in this sense there are no local ({\it off-shell}) observables in gravity.

Let us now try to quantify the amount of indetermination due
to quantum gravity. The parameter controlling the uncertainty $1/S=G/r^{D-2}$
is always tiny for distances larger than the Planck length,
which signals the fact that quantum gravity becomes important 
at this scale. We do not expect the low-energy effective 
theory to break down at any order in perturbation theory,
{\it i.e.} at any order in $1/S$. This is what perturbative 
string theory suggests by providing, in principle, a well defined 
higher-derivative low-energy expansion at all order in $G$.
Also, in our 2-point function example, 
the natural limit on the resolution should be set by
the number of states of the apparatus ($e^S$) instead of its number of d.o.f. ($S$).
We thus expect the irreducible error due to quantum gravity to be non-perturbative in 
the coupling $G$,
\beq
\delta_{ \la \phi(x) \phi(y)\ra }\sim e^{-S}  \sim e^{ -\frac{(x-y)}{G}^{D-2}}
\eeq
The smallness and the non-perturbative nature of this effect 
suggest that it becomes important only at very short distances, with
the low-energy field theory remaining a very good approximation
at long distances. This is true except in special situations 
where the effective theory breaks down when
it is not naively expected to. However, before discussing this point further, 
let us examine the issue of locality from another angle by looking at what it means in $S$-matrix language. 

As is well-known, the $S$-matrix associated 
with a local theory enjoys analyticity properties.
For instance, for the $2 \to 2$ scattering, the amplitude must be an
analytic function of the Maldelstam's variables $s$ and $t$ away from the real axis. It must also be
exponentially bounded in energy---at fixed angles, the amplitude can
not fall {\it faster} than $e^{-\sqrt{s} \log s}$~\cite{expbound}. In local QFT, both of these
requirements follow directly from the sharp vanishing of field
commutators outside the light-cone in position space. A trivial
example illustrates the point: consider a function $f(x)$ that
vanishes sharply outside the interval $[x_1,x_2]$. What does this
imply for the Fourier transform $\tilde{f}(p)$? Since the integral
for $\tilde{f}(p)$ is over a finite range $[x_1,x_2]$ and $e^{i p x}$
is analytic in $p$, $\tilde{f}(p)$ must be both analytic and
exponentially bounded in the complex $p$ plane.
Now amplitudes in UV complete local quantum field theories certainly
satisfy these requirements---they are analytic and fall off as powers
of energy. More significantly, amplitudes in perturbative string
theory {\it also} satisfy these bounds. That they are analytic is no
surprise, since after all the Veneziano amplitude arose in the
context of the analytic S-matrix program. More non-trivially they
are also exponentially bounded---high energy amplitudes for $E \gg
M_s$ are dominated by genus $g = E/M_s$ and fall off precisely as
$e^{-E \log E}$, saturating the locality bound~\cite{MO} (see also \cite{Vene} and referenced therein for discussion of high-energy scattering 
in string theory). Thus despite
naive appearances, the finite extent of the string does not in
itself give rise to any violations of locality. Indeed, we now know
of non-gravitational string theories---little string theories in six
dimensions. These theories have a definition in terms of
four-dimensional gauge theories via deconstruction and are
manifestly local in this sense~\cite{6d-deconstruction}.

It is possible that violations of locality do show up in the S-matrix when black hole production becomes important. 
At high enough energies relative to the Planck scale,  
the two-particle scattering is dominated by black hole production, when the energy becomes larger than $M_{Pl}$ divided by 
some power of $g_s$ so the would-be BH becomes larger than the string scale. The $2 \to 2$
scattering amplitude therefore cannot be smaller than $e^{-S(E)}$, and it is natural to conjecture that this lower bound is met:
\beq \label{eq:scatt}
{\cal A}_{2\to 2}(E \ggg M_{\rm Pl}) \sim e^{-S(E)} \sim e^{-E R(E)}
\eeq 
where $R(E) \sim (G E)^{1/(D-3)}$ is the radius of the black hole 
formed with center of mass energy $E$ and $S(E)$ is the associated entropy. Note that since $R(E)$ grows as a power of energy, saturating this lower bound leads to an amplitude falling faster than exponentially at high energies, so that the only sharp mirror of locality in the scattering amplitude is lost. 
A heuristic measure of the size of these non-local effects in position space can be 
obtained by Fourier transforming the analytically continued ${\cal A}_{2 \to 2}$ back to 
position space; a saddle point approximation using the black-hole dominated amplitude 
gives a Fourier transform of order $e^{-r^{D-2}/G} \sim e^{-S}$, in accordance with our general expectations. Of course this asymptotic form of the scattering amplitude is a guess; it is hard to imagine that the amplitude is smaller than this but one might imagine that it can be larger (we thank J. Maldacena for pointing this out to us). The point is that there is no reason to expect perturbative string effects to violate notions of locality---they certainly do not in the S-matrix---while gravitational effects can plausibly do it. 

Naively one would expect that the breakdown of locality only shows up when scales of order of the Planck length
or shorter are probed, while for IR physics
the corrections are ridiculously tiny ($e^{-S}$) with no observable effects.
This is however not true. There are several important cases where the loss
of locality by quantum gravity give ${\cal O}(1)$ effects.
This happens when in processes with ${\cal O}(e^S)$ states, the tiny ${\cal O}(e^{-S})$ 
corrections sum to give ${\cal O}(1)$ effects.
This is similar to renormalon contributions in QCD. Independently
of the value of $\alpha_s$, or equivalently of the energy considered,
every QCD amplitude is indeed affected by non-perturbative power corrections
\beq
\frac{\Lambda^2_{{\rm QCD}}}{Q^2}\sim e^{-\frac{1}{\beta_0\, \alpha_s\left ( Q^2 \right )}}
\eeq
which limit the power of the ``asymptotic" perturbative expansion.
Because in the $N$-loop order contributions, and equivalently in the $N$-point functions,
combinatorics produce enhancing $N!$ factors, they start receiving 
${\cal O}(1)$ corrections when $N\sim 1/\alpha_s$.
Analogously in gravity, we must expect ${\cal O}(1)$ corrections 
from ``non-perturbative" quantum gravity in processes with $N$-point functions
with $N\simeq S$. These contributions are not captured by the perturbative expansion,
they show the very nature of quantum gravity  and its non-locality,
which is usually thought to be confined at the Planck scale. 
Indeed in eq.~(\ref{eq:scatt}) it is the presence of $e^S$ states (the inclusive amplitude is an $S$-point function) 
that suppresses exponentially the $2\to2$ amplitude, thus violating the locality bound.
An example where this effect becomes macroscopic
is well-known as the black hole information paradox~\cite{Hawk-bhip},
and will be reviewed more extensively below in section~\ref{sec:BH}.
Notice however that only for specific questions $e^{-S}$ effects become relevant,
in all other cases, where less than ${\cal O}(S)$ quanta are involved,
the low energy effective theory of gravity (or perturbative string theory)
remains an excellent tool for describing gravity at large distances.



\subsection{The black hole information paradox}\label{sec:BH}

Since an effective field theory analysis of black hole information and evaporation leads to dramatically incorrect conclusions, it is worth reviewing this well-worn territory in some detail, in order to draw a lesson that can then be applied to cosmology. 

Schwarzschild black hole solutions of mass $M$ and radius $R_S$ (with $R_S^{D-3} \sim G M$) exist for any $D > 3$ spacetime dimensions. Black holes lose mass via Hawking radiation~\cite{Hawking:1974sw} 
with a rate $dM/dt \sim - R^{-2}$, so that the evaporation time is
\begin{equation}
\frac{t_{\rm ev}}{R_S} \sim M R_S \sim S_{\rm BH}
\end{equation}
where
\begin{equation}
\label{BHentropy}
S_{\rm BH} \sim \frac{R_S^{D-2}}{G}
\end{equation}
is the black hole entropy, the large dimensionless parameter in the
problem. Note that there is a natural limit where the geometry ($R_S$)
is kept fixed, $M\to\infty$ but $G \to 0$ so that $S_{\rm BH} \to \infty$. In this
limit there is still a black hole together with its horizon and
singularity, and it emits Hawking radiation with temperature $T_H\sim R_S^{-1}$, but $t_{\rm ev} \to \infty$
so the black hole never evaporates.

Hawking radiation can certainly be computed using effective field theory, after all the horizon of a macroscopic black hole is a region of spacetime with very small curvature and as a consequence there should be a description of the evaporation where only low-energy degrees of freedom are excited. 
In order to derive Hawking radiation, one has to be able to describe the evolution of an initial state on the black hole semiclassical background to some final state that has Hawking quanta. Following the laws of quantum mechanics, all that is needed is a set of spatial slices and the corresponding---in general time dependent---Hamiltonian. 
\begin{figure}[t!]
\begin{center}
\includegraphics[height=7cm]{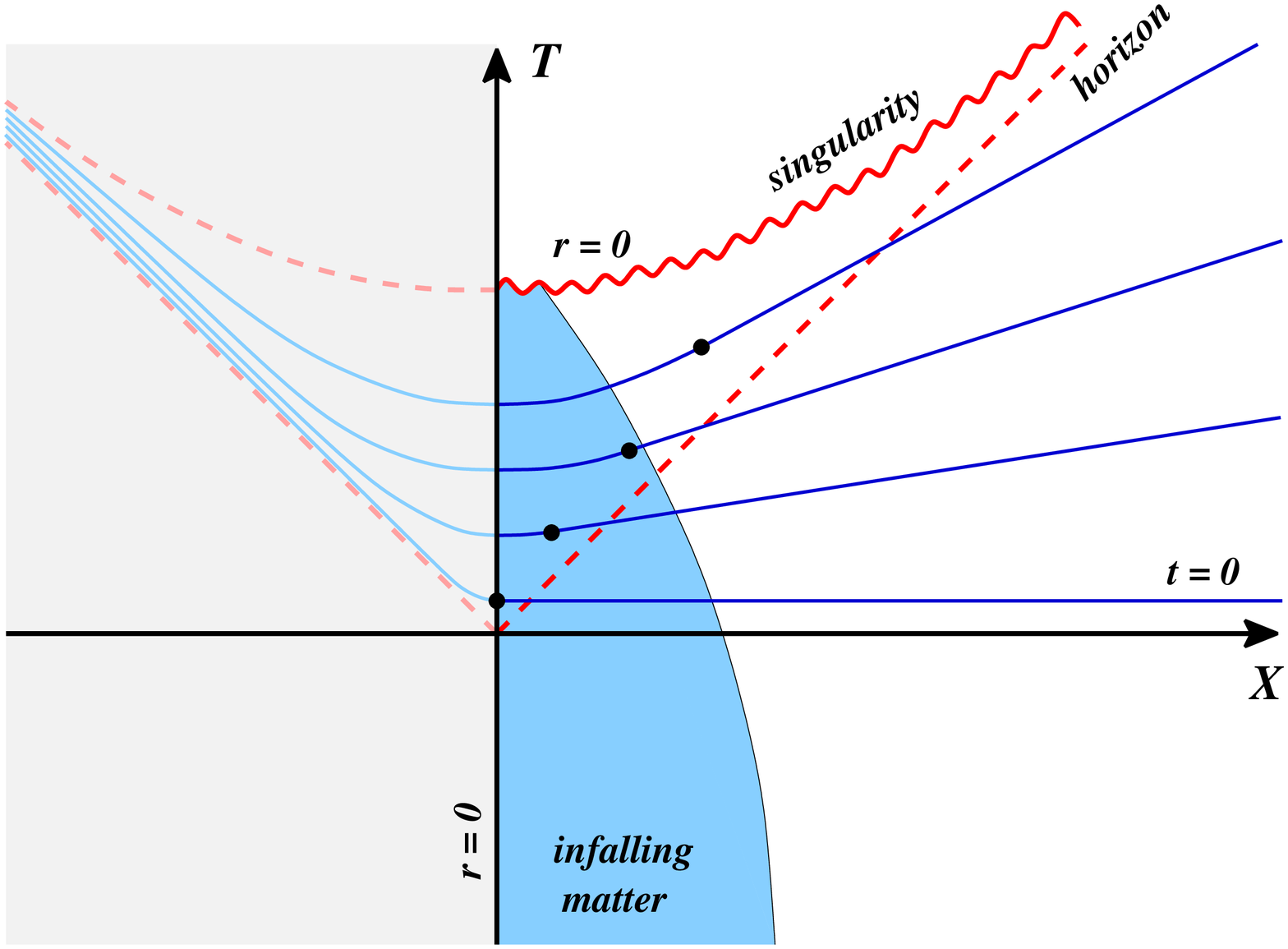}
\hspace{2cm}
\includegraphics[height=7cm]{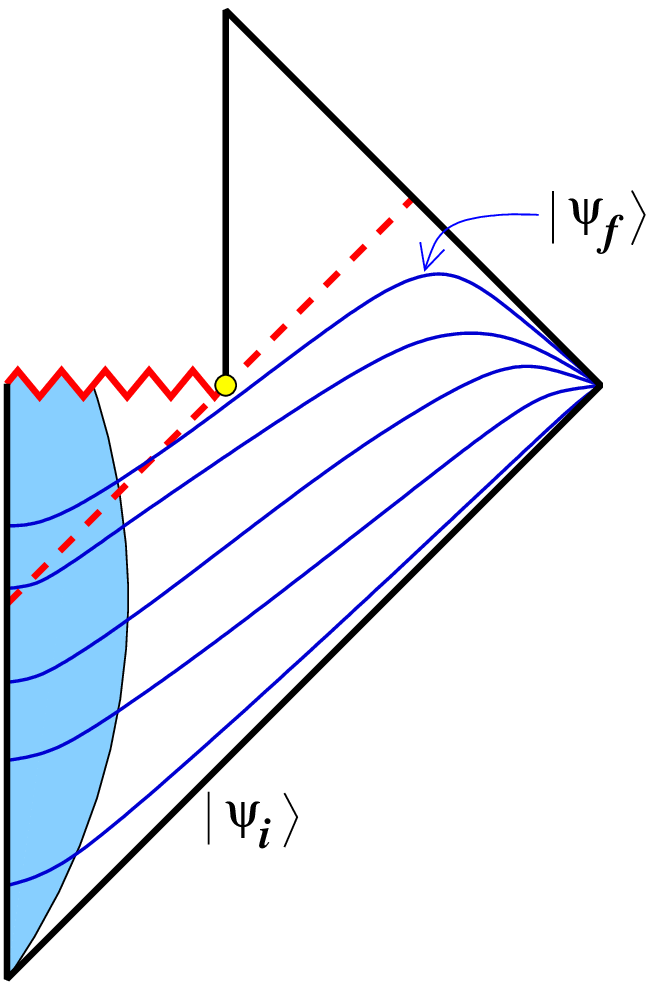}
\caption{\label{niceslices_fig} \small \it Nice slices in Kruskal coordinates {\em (left)} and in the Penrose diagram {\em (right)}. The singularity is at $T^2-X^2=1$.}
\end{center}
\end{figure}
However, because the aim is to compute the final state within a long distance effective field theory, 
the curvature of the sliced region of spacetime must be low everywhere (the slices can also cross the horizon if they stay away from the singularity) and the extrinsic curvature of the slices themselves has to be small as well. Spatial surfaces with these properties are called ``nice slices" \cite{WaldNice,Polchinski:1995ta}. 
One can easily arrange for this slicing to cover also most of the collapsing matter that forms the black hole. To be specific we can take the first ($t=0$) slice to be $T= c_0$  for $X\geqslant 0$ and the hyperbola $T^2-X^2= c_0^2$ for $X<0$, where $X$ and $T$ are Kruskal coordinates; this slice has small extrinsic curvature by construction. Then we take a second slice with $c_1>c_0$ and we boost it in such a way that the asymptotic Schwarzschild time on this slice is larger than the asymptotic time on the previous one. We can build in this way a whole set of slices $c_0 < \ldots< c_n$, all with small extrinsic curvature; if $c_n \lesssim  \frac{1}{2}$ the region they cover inside the horizon is still far away from the singularity, while outside they can be boosted arbitrarily far in the future (Fig.~\ref{niceslices_fig}) so that they can intercept most of the outgoing Hawking quanta. When the black hole evaporates the background geometry changes and the slices can be smoothly adjusted with the change in the geometry until very late in the evaporation process, when the curvature becomes Planckian and the black hole has lost most of its mass.

Starting with a pure state $| \psi_i \rangle$ at $t=0$, one can now evolve it using the Hamiltonian $H_{\rm NS}$ defined on this set of slices, never entering the regime of high curvature.
We can now imagine dividing the slices in a portion that is outside the horizon and one inside it; even if the state on the entire slice is pure, we can consider the effective density operator outside the black hole defined as $\rho_{\rm out}(t)=$ Tr$_{\rm in}$ $|\psi (t)\rangle \langle \psi(t) |$. In principle we can measure $\rho_{\rm out}$. As usual in quantum mechanics, this is done by repeating exactly the same experiments an infinite number of times, and measuring all the mutually commuting observables that are possible. We should certainly expect that at early times $\rho_{\rm out}$ is a mixed state, representing the entanglement between infalling matter and Hawking radiation along the early nice-slices. 
This can be quantified by looking at the {\it entanglement entropy} associated with $\rho_{\rm out}$:
\be
S_{\rm ent} = - \textrm{Tr} \;  \rho_{\rm out} \log \rho_{\rm out}
\ee
Clearly at early times $S_{\rm ent}$ is non-vanishing. What happens at late times? 
Should we expect the final state of the evolution to be $|\psi_f \rangle = | \psi_{\rm out} \rangle \otimes |\psi_{\rm in}\rangle$,  with no entanglement between inside and outside and $S_{\rm ent}=0$? The answer is negative because of the quantum Xerox principle \cite{Bigatti:1999dp}. If this decomposition were correct, two different states $| A \rangle$ and $| B \rangle$ should evolve into
\be
| A \rangle   \rightarrow | A_{\rm out} \rangle \otimes |A_{\rm in} \rangle , \; \qquad | B \rangle \rightarrow | B_{\rm out} \rangle \otimes |B_{\rm in} \rangle
\ee  
but a linear superposition of them
\be
\sfrac{1}{\sqrt{2}} \big( | A \rangle + | B \rangle \big) \ \rightarrow \ \sfrac{1}{\sqrt{2}} \big( | A_{\rm out} \rangle \otimes |A_{\rm in} \rangle +  | B_{\rm out} \rangle \otimes |B_{\rm in} \rangle \big) 
\ee  
cannot be of the form $ (| A \rangle + | B \rangle )_{\rm out} \otimes ( | A \rangle + | B \rangle)_{\rm in} $ unless the states behind the horizon are equal $| A_{\rm in} \rangle = | B_{\rm in} \rangle$ for every $A$ and $B$, and this is clearly impossible.
No mystery then that the outgoing Hawking radiation $\rho_{\rm out}$ looks thermal, being correlated with states behind the horizon.

\begin{figure}[t!]
\begin{center}
\includegraphics[width=10cm]{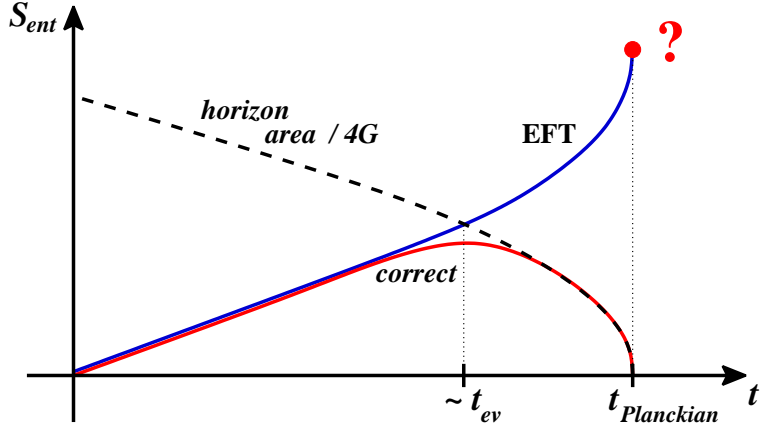}
\caption{\label{entang_fig} \small \it The entanglement entropy for an evaporating black hole as a function of time. After a time of order of the evaporation time the EFT prediction {\em (blue line)} starts violating the holographic bound {\em (dashed line)}.
The correct behavior {\em (red line)} must reduce to the former at early times and approach the latter at late times. At the final stages, $t \simeq t_{\rm Planckian}$,
curvatures are large and EFT breaks down.}
\end{center}
\end{figure}

Using nice slices one can compute in the low energy theory the entanglement entropy associated with the density matrix $\rho_{\rm out}$: while the horizon area shrinks, the number of emitted quanta increases, 
the entanglement entropy of these thermal states grows monotonically as a function of time until the black hole becomes Planckian, the effective field theory is no longer valid and we don't know what happens next without a UV completion (Fig.~\ref{entang_fig}).
This seems a generic prediction of low energy EFT. It implies a peculiar fate for black hole evaporation: either the evolution of a pure state ends in a mixed state, violating unitarity, or the black hole doesn't evaporate completely, a Planckian remnant is left and the information remains stored in the correlations between Hawking radiation and the remnant. What cannot be is that the purity of the final state is recovered in the last moments of black hole evaporation, because the number of remaining quanta is not large enough to carry all the information. This is the black hole information paradox.
It suggests that in order to preserve unitarity, effective field theory should break down earlier than expected. If we believe in the holographic principle, the total dimension of the Hilbert space of the region inside the black hole has to be bounded by the exponential of the horizon area in Planckian units. Since the entropy of any density operator is always smaller than the logarithm of the dimension of the Hilbert space, and since the entanglement entropy for a pure state divided into two subsystems is the same for each of them,  the correct value of the entanglement entropy that is measured from $\rho_{\rm out}$ should start decreasing at a time of order $t_{\rm ev}$, finally becoming zero when the black hole evaporates and a pure state is recovered.

According to this picture, the difference between the prediction of EFT and the right answer is of $\mathcal{O}(1)$ in a regime where curvature is low and there is no reason why effective field theory should be breaking down.
However, the way this $\mathcal{O}(1)$ difference manifests itself is rather subtle. To understand this point let us first consider  $N$ spins $\sigma_i = \pm \frac{1}{2}$ and take the following state:
\be
|\psi \rangle = \sum_{\{ \sigma_i\}} \frac{1}{2^{N/2}} |\sigma_1\ldots \sigma_N \rangle e^{i \theta(\sigma_1,\ldots,\sigma_N)}
\ee
where $\theta(\sigma_1,...,\sigma_N)$ are random phases. If only $k$ of the $N$ spins are measured, the density matrix $\rho_k$ can be computed taking the trace over the remaining $N-k$ spins
\be
\label{spinmat}
\rho_k = \frac{1}{2^k}  \sum |\sigma_1 \ldots \sigma_k \rangle \langle \sigma_1 \ldots \sigma_k | + \mathcal{O} (2^{-\frac{N+k}{2}})_{\textrm{off-diagonal}} 
\ee
the off-diagonal exponential suppression comes from averaging $2^{N-k}$ random phases. 
When $k \ll N$ this density matrix looks diagonal and maximally mixed. Let us now study the entanglement entropy: for small $k$ we can expand 
\be
S_{ent} = - \textrm{Tr} \;  \rho_{k} \log \rho_{k} = k \log{2} + \mathcal{O}(2^{-N+2k}) 
\ee
and conclude that the effect of correlations becomes important only when $k \simeq N/2$ spins are measured; finally when $k\sim N$ the entanglement entropy goes to zero as expected for a pure state. A state that looks thermal instead of maximally mixed is  
$| \psi \rangle = \sum_{E_n} e^{-\beta \frac{E_n}{2}} |E_n \rangle e^{i \theta(E_n)}$ with random phases $\theta(E_n)$.  This is of course why common pure states in nature, like the 
proverbial ``lump of coal" entangled with the photons it has emitted, look thermal when only a subset of the states is observed. 
 
This is a simple illustration of a general result due to Page \cite{Page:1993wv}, showing how the difference between a pure and a mixed state is exponentially small until a number of states of order of the dimensionality of the Hilbert space is measured. 
Suppose we have to verify if the black hole density operator has an entanglement entropy of order $S$, then we need to measure an $e^S \times e^S$ matrix---the entropy of any $N\times N$ matrix is bounded by $\log N$---with entries of order $e^{-S}$; in order to see $\mathcal{O}(1)$ deviations from thermality in the spectrum, a huge number of Hawking states must be measured with incredibly fine accuracy.   
Because it takes a time scale of order of the evaporation time $t_{\rm ev}= R_S S_{\rm BH}$ to emit order $S_{\rm BH}$ quanta, before that time effective field theory predictions are correct up to tiny $e^{-S}$ effects (Fig.~\ref{entang_fig}).
In particular this means that when looking at the $N$-point functions of the theory, the exact value is the one obtained using EFT plus corrections that are exponentially small until $N\simeq S$:
\be
\langle \phi_1 \ldots \phi_N \rangle_{\textrm{correct}} = \langle \phi_1 \ldots \phi_N \rangle_{\textrm{EFT}} + \mathcal{O}(e^{-(S-N)})
\ee
This can be explicitly seen with large black holes in AdS, as discussed by Maldacena \cite{Maldacena:2001kr} and Hawking \cite{Hawking:2005kf}. The semiclassical boundary two-point function for a massless scalar field falls off as $e^{-t/R}$. Its vanishing as $t \to \infty$ is the information paradox in this context, while the CFT ensures that this two-point function never drops below $e^{-S}$; but the discrepancy of the semiclassical approximation relative to the exact unitary CFT result for the two-point function is of order $e^{-S}$. 

There is another heuristic observation that supports the idea that the whole process of black hole evaporation cannot be described within a single effective field theory. There is actually a limitation in the slicing procedure that we described at the beginning of this section. In order for the slices to extend arbitrarily in the future outside the black hole, they have to be closer and closer inside the horizon. However, quantum mechanics plus gravitation put a strong constraint: to measure shorter time intervals  heavier clocks are needed. Of course they must be larger than their own Schwarzschild radius but a clock has also to be smaller than the black hole itself.
This gives a bound on the shortest interval of time $\delta t$ (the difference $c_k-c_{k-1}$ between two subsequent slices in (Fig.~\ref{niceslices_fig})) that makes sense to talk about inside the horizon
\be
\delta t \gtrsim \frac{\hbar}{M_{\rm clock}} \gtrsim \frac{\hbar G}{R_S^{D-3}}
\ee
In this equation we have temporarily restored $\hbar$ to highlight the fact that whenever $\hbar$ or $G$ goes to zero the bound becomes trivial.
On the other hand, the proper time inside the black hole is finite  $\tau_{\rm in}\lesssim
R_S$. These two conditions imply a striking bound: the maximum number of slices inside the black hole is also finite, $N_{\rm max}\simeq \tau_{\rm in}/\delta t \simeq R_S^{D-2}/G \simeq S_{\rm BH}$.
How large is then the time interval that we can cover outside? With a spacing between the slices of the order of the Planck length ($\ell_{\rm Pl}$) the total time interval is $\tau_{\rm out}\simeq N_{\rm max} \ell_{\rm Pl} \simeq R_S^{D-2} G^{(3-D)/(D-2)}$. Note, however, that if we are only interested in the Hawking quanta we may allow for a much less dense slicing: the spacing outside can be of order of the typical wavelength of the radiation $\delta t_{\rm out} \sim 1/T_{\rm BH} \sim R_S$. In this way we can cover at most 
\be
\tau_{\rm out}\lesssim N_{\rm max} R_S \simeq S_{\rm BH} R_S
\ee 
which is precisely the evaporation time $t_{\rm ev}$. Summarizing, the system of slices we need to define the Hamiltonian evolution cannot cover enough space-time to describe the process of black hole evaporation for time intervals parametrically larger than $t_{\rm ev}$. 
With this argument we find that effective field theory should break down exactly when it starts giving the wrong prediction for the entanglement entropy (Fig.~\ref{entang_fig}). Most previous estimates instead accounted for a much shorter regime of validity, up to time-scales of order $R_S \log R_S$ \cite{Lowe:1995ac,Giddings:2004ud}. This would imply that the EFT breakdown originates at some finite order in perturbation theory while in our case it comes from non-perturbative $\mathcal{O}(e^{-S})$ effects.

Because the EFT description of states is incorrect for late times $t \gg t_{\rm ev}$, also the association of commuting operators to the 2-point function is wrong. It can well be that two observables evaluated on the same nice slice, one inside the horizon and the other outside, will no longer commute, even if they are at large spatial separation. This is consistent with the principle of black hole complementarity \cite{'tHooft:1990fr,Susskind:1993if}. Notice however that this breaking down is not merely a kinematical effect due to the presence of the horizon, after all EFT is perfectly good for computing Hawking radiation. In fact in the limit described at the beginning of this section, when we keep the geometry fixed and we decouple dynamical gravity, there still is an horizon but effective field theory now gives the right answer for arbitrarily long time scale: the black hole doesn't evaporate and information is entangled with states behind the horizon. 
The limitation on the validity of EFT comes from  {\it dynamical} gravity. There is nothing wrong in talking about both the inside and the outside of the horizon for time intervals parametrically smaller than $t_{\rm ev}$ and even if one goes past that point, $\mathcal{O}(S)$ quanta have to be measured to see a deviation of $\mathcal{O}(1)$.

 
\section{Limits on de~Sitter space}\label{sec:dS}
We now consider de~Sitter space. According to the covariant entropy bound, de~Sitter space should have a finite maximum entropy given in 4D by the horizon area in $4G$ units, $S_{\rm dS} = \pi H^{-2}/G$. For a black hole in asymptotically flat space it makes sense that the number of internal quantum 
states should be finite. After all for an external observer a black hole is a localized object, occupying a limited region of space. But for de~Sitter space it is less clear how to think about the finiteness of the number of quantum states: de~Sitter has infinitely large spatial sections, at least in flat FRW slicing, and continuous non-compact isometries---features that seem to clash with the idea of a finite-dimensional Hilbert space.
In particular the de~Sitter symmetry group $SO(n,1)$ has no finite-dimensional representations, so it cannot be realized in the de~Sitter Hilbert space 
(see however Ref.~\cite{witten} for a discussion on this point).
However the fact that no single observer can ever experience what is beyond his or her causal horizon makes it tempting
to postulate some sort of  `complementarity' between the outside and the inside of the horizon, in the same spirit as the black hole complementarity. From this point of view the global picture of de~Sitter space would not make much sense at the quantum level.

It is plausible that the global picture of de~Sitter space is only a semiclassical approximation, which becomes strictly valid only in the limit where gravity is decoupled while the geometry is kept fixed. In the same limit the entropy $S_{\rm dS}$ diverges, and one recovers the infinite-dimensional Hilbert space of a local QFT in a fixed de~Sitter geometry.
With dynamical gravity we expect tiny non-perturbative effects of order $e^{-S_{\rm dS} }$ to put fundamental limitations on how sharply one can define local observables, in the spirit of sect.~\ref{sec:locality}. These tiny effects can have dramatic consequences in situations where they are enhanced by huge $\sim e^{+S_{\rm dS}}$ multiplicative factors. For instance it is widely believed that on a timescale of order $H^{-1} e^{S_{\rm dS}}$---the Poincar\'e recurrence time---de~Sitter space necessarily suffers from instabilities and no consistent theory of pure de~Sitter space is possible; although this view has been seriously challenged by Banks \cite{Banks:2002nm,Banks:2004xh,Banks:2005bm}.  Notice
however that the near-horizon geometry of de~Sitter space is identical to that of a black hole---they are both equivalent to Rindler space. 
As we discussed in sect.~\ref{sec:BH}, in the black hole case the local EFT description must break down at a time $t_{\rm ev} \sim R_S \cdot S_{\rm BH}$ after the formation of the black hole itself.  It is natural to conjecture that a similar breakdown of EFT occurs in de~Sitter space after a time of order $H^{-1} \cdot S_{\rm dS}$. This is an extremely shorter timescale than the Poincar\'e recurrence time, which is instead exponential in the de~Sitter entropy.

\subsection{Slow-roll inflation}
Is there a way to be more concrete?
In pure de~Sitter any observer has access only to a small portion of the full spacetime, and it is not even clear what the observables are \cite{bousso}.
But we can make better sense of de~Sitter space if we regulate it by making it a part of inflation. If inflation eventually ends in a flat FRW cosmology with zero cosmological constant, then asymptotically in the future every observer will have access to the whole of spacetime. In particular an asymptotic observer can detect---in the form of density perturbations---modes that exited the cosmological horizon during the near-de~Sitter inflationary epoch. Notice that from this point of view it looks perfectly sensible to talk about what is outside the early de~Sitter horizon---we even have {\em experimental} evidence that computing density perturbations by following quantum fluctuations outside the horizon is reliable---and a strict complementarity between the inside and the outside of the de~Sitter horizon seems too restrictive.
Now, the interesting point is that the fact that an asymptotic observer can detect modes coming from the early inflationary phase gives an operational meaning to the de~Sitter degrees of freedom, and to their number. Every detectable mode corresponds to a state in the de~Sitter Hilbert space. 

Let's consider for instance an early phase of ordinary slow-roll inflation. Classically the inflaton $\phi$ rolls down its potential $V(\phi)$ with a small velocity $\dot \phi_{\rm cl}\sim V'/H$.
On top of this classical motion there are small quantum fluctuations. Modes get continuously stretched out of the de~Sitter horizon, and quantum fluctuations get frozen at their typical amplitude at horizon-crossing,
\be \label{deltaphi}
\delta \phi_{\rm q} \sim H 
\ee
For a future observer, who makes observations in an epoch when the inflaton is no longer an important degree of freedom, these fluctuations are just small fluctuations of  the space-like hyper-surface that determines the end of inflation. That is, since with good approximation inflation ends at some fixed value of $\phi$, small fluctuations in $\phi$
curve this hypersurface by perturbing the local scale factor $a$,
\be \label{deltaa}
\frac{\delta a}{a} \sim H \delta t \sim H \, \frac{\delta \phi_{\rm q}}{\dot \phi_{\rm cl}} 
\ee
Such a perturbation is locally unobservable as long as its wavelength is larger than the cosmological horizon. But eventually every mode re-enters the horizon, and when this happens a perturbation in the local $a$ translates into a perturbation in the local energy density $\rho$,
\be \label{deltarho}
\frac{\delta \rho}{\rho} \sim \frac{\delta a}{a} \sim \frac{H^2}{\dot \phi_{\rm cl}} 
\ee
where we made use of eq.~(\ref{deltaphi}).

By observing density perturbations in the sky an asymptotic observer is able to assign states to the approximately de~Sitter early phase.
If we believe the finiteness of de~Sitter entropy, the maximum number of independent modes from inflation an observer can ever detect should be bounded by the dimensionality of the de~Sitter Hilbert space, ${\rm dim}({\cal H}) = e^{S}$. 
Of course slow-roll inflation has a finite duration, thus only a finite number of modes
can exit the horizon during inflation and re-enter in the asymptotic future. Roughly speaking, if inflation lasts for a total of $N_{\rm tot}$ $e$-foldings, the number of independent modes coming from inflation is of order $e^{ 3 N_{\rm tot}}$---it is the number of different Hubble volumes that get populated starting from a single inflationary Hubble patch.
If the number of $e$-foldings during inflation gets larger than the de~Sitter entropy, $N_{\rm tot} \gtrsim S$, this operational definition of de~Sitter degrees of freedom starts violating the entropy bound.

In slow-roll inflation the Hubble rate slowly changes with time, 
\be  \label{Friedmann}
\dot H = - (4 \pi G) \, \dot \phi^2  
\ee
and so does the associated de~Sitter entropy $S = \pi H^{-2}/G$. In particular, the rate of entropy change per $e$-folding is
\be
\frac{dS}{dN} = \frac{8 \pi^2 \dot \phi^2  }{H^4} 
\sim \bigg(\frac{\delta \rho}{\rho} \bigg)^{-2}
\ee
where we made use of eq.~(\ref{deltarho}). By integrating this equation we get a bound on the total number of $e$-foldings,
\be \label{Ntot}
N_{\rm tot} \lesssim \bigg(\frac{\delta \rho}{\rho} \bigg)^{2} \cdot S_{\rm end} 
\ee
where $S_{\rm end}$ is the de~Sitter entropy at the end of inflation. We thus see that since $\delta \rho/ \rho$ is smaller than one, the total number of $e$-foldings is bounded by the de~Sitter entropy. As a consequence a future observer will never be able to associate more than $e^S$ states to the near-de~Sitter early phase!

By adjusting the model parameters one can make the inflationary potential flatter and flatter, thus enhancing the amplitude of density perturbations $\delta \rho/\rho$. In this way, according to eq.~(\ref{Ntot}) for a fixed de~Sitter entropy the allowed number of $e$-foldings can be made larger and larger. When $\delta \rho / \rho$ becomes of order one we start saturating the de~Sitter entropy bound, $N_{\rm tot} \sim S$. However exactly when $\delta \rho/ \rho$ is of order one we enter the regime of eternal inflation. Indeed quantum fluctuations in the inflaton field, $\delta \phi_{\rm q} \sim H$, are so large that they are of the same order as the classical advancement of the inflaton itself in one Hubble time, $\Delta \phi_{\rm cl} \sim \dot \phi_{\rm cl} \cdot H^{-1}$,
\be
\frac{\delta \phi_{\rm q} }{\Delta \phi_{\rm cl}} \sim \frac{\delta \rho}{\rho} \sim 1
\ee
Now in principle there is no limit to the total number of $e$-foldings one can have in an inflationary patch---the field can fluctuate {\em up} the potential as easily as it is classically rolling down. Still when a future observer starts detecting
modes coming from an eternal-inflation phase, precisely because they correspond to density perturbations of order unity the Hubble volume surrounding the observer will soon get collapsed into a black hole~\cite{Linde, bousso_inside}. Therefore a future observer will not be able to assign more than $e^S$ states to the inflationary phase.

Notice that when dealing with eternal inflation we are pushing the semiclassical analysis beyond its regime of validity, by applying it to a regime of large quantum fluctuations. This is to be contrasted with standard (i.e., non-eternal) slow-roll inflation, where the semiclassical computation is under control and quantitatively reliable.
This matches nicely with what we postulated above by analogy with the black hole system---that in de~Sitter space the local EFT description should break down after a time of order $H^{-1} \cdot S$. Indeed 
in standard slow-roll inflation the near-de~Sitter phase cannot be kept for longer than $N_{\rm tot} \sim S$ $e$-foldings. 

Normally whether inflation is eternal or not is controlled by the microscopic parameters of the inflaton potential. For slow-roll inflation we have just given instead a {\em macroscopic} characterization of eternal inflation, involving geometric quantities only: an observer living in an inflationary Universe can in principle measure the local $H$ and $\dot H$ with good accuracy, and determine the rate of entropy change per $e$-folding. If such a quantity is of order one, the observer lives in an eternally inflating Universe.

Indeed we will see that this macroscopic characterization of eternal inflation 
is far more general than the simple single-field slow-roll inflationary model we are discussing here. By now we know several alternative mechanisms for driving inflation, well known examples being for instance DBI inflation \cite{DBI}, locked inflation \cite{locked}, k-inflation \cite{k}. These models can be thought of as different regularizations of de~Sitter space---different ways of sustaining an approximately de~Sitter early phase for a finite period of time before matching onto an ordinary flat FRW cosmology, thus allowing an asymptotic observer to gather information about de~Sitter space.
We will show in a model-independent fashion that the absence of eternal inflation requires that the Hubble rate decrease faster than a critical speed, 
\be \label{Hdot}
|\dot H| \gg G H^4 
\ee
This is a {\em necessary} condition for the classical motion not to be overwhelmed by quantum fluctuations, so that the semiclassical analysis is trustworthy. 
In terms of the de~Sitter entropy the above inequality reads
\be \label{dS}
\frac{dS}{dN} \gg 1 
\ee
which once integrated limits the total number of $e$-folds an inflationary model can achieve without entering an eternal-inflation regime,
\be
N_{\rm tot} \ll S_{\rm end} 
\ee

As pointed out by Bousso, Freivogel and Yang, the bound (\ref{dS}) is necessarily violated \cite{bousso_inside} in slow-roll eternal inflation,  thereby avoiding conflict with the second law of thermodynamics. Indeed, during eternal inflation the evolution of the horizon area is dominated by quantum jumps of the inflaton field and can go either way during each $e$-folding. From $\left|\frac{dS}{dN} \right|<1$ one infers that the entropy changes by less than one unit during each $e$-folding and, consequently, its decrease is unobservable. 

One notable exception is ghost inflation \cite{ghost_I}. There $\dot \phi$ and $\dot H$ are not tightly bound to each other like in eq.~(\ref{Friedmann}). Indeed there exists an exactly de~Sitter solution with vanishing $\dot H$ but constant, non-vanishing $\dot \phi$. This is because the stress-energy tensor of the ghost condensate vacuum is that of a cosmological constant, even though the vacuum itself breaks Lorentz invariance through a non-zero order parameter $\langle \dot \phi \rangle$ \cite{ghost_C}. Therefore, the requirement of not being eternally inflating still gives a lower bound on $\dot \phi$
but now this does not translate into a lower bound on $|\dot H|$. $\dot H$ can be strictly zero, still inflation is guaranteed to end by the incessant progression of the scalar, which will eventually trigger a sudden drop in the cosmological constant \cite{ghost_I}. Thus in ghost inflation there is no analogue of the local bounds (\ref{Hdot}) and (\ref{dS}), nor there is any upper bound on the total number of $e$-foldings.

Notice however that the ghost condensate is on the verge of violating the null energy condition, having $\rho+p = 0$. Indeed small perturbations about the condensate do violate it.
In the next subsection we will prove that our bounds are guaranteed to hold for all inflationary systems that do not admit violations of the null energy condition. 
This matches with the general discussion of sect.~\ref{sec:NEC}: the NEC is known to play an important role in the holographic bound and in general in limiting the accuracy with which one can define local observables in gravity. The fact that  all reliable NEC-respecting semiclassical models of inflation obey our bounds, suggests that the latter really limit the portion of de~Sitter space one can consistently talk about within local EFT.

\begin{figure}[t!]
\begin{center}
\includegraphics[width=8cm]{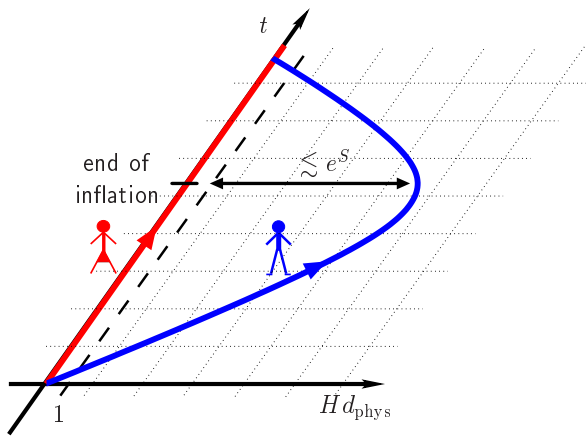}
\caption{\small \it Love in an inflationary Universe.}
\end{center}
\end{figure}

\subsection{General case}\label{sec:proof}

Let us consider a generic inflationary cosmology driven by a collection of matter fields $\psi_m$. We want to see under what conditions the time-evolution of the system is mainly classical, with quantum fluctuations giving only negligible corrections. We could work with a completely generic matter Lagrangian, function of the matter fields and their first derivatives, and possibly including higher-derivative terms, which in specific models like ghost inflation can play a significant role. We should then: take the proper derivatives with respect to the metric to find the stress-energy tensor; plug it into the Friedmann equations and solve them; expand the action at quadratic order in the fluctuations around the classical solution; compute the size of typical quantum fluctuations; impose that they do not overcome the classical evolution. This procedure would be quite cumbersome, at the very least.

\begin{figure}[t!]
\begin{center}
\includegraphics[width=14cm]{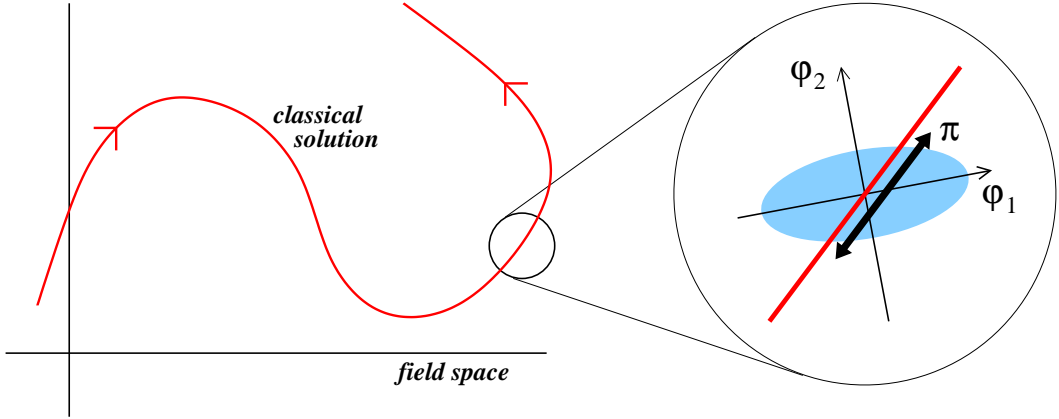}
\caption{\small \it \label{pi_figure} A given cosmological history is a
classical trajectory in field space (red line), parameterized by
time. The Goldstone field $\pi$ describes small local fluctuations
{\em along} the classical solution. In general other light
oscillation modes, transverse to the trajectory will also be
present, and $\pi$ can be mixed with them. In the picture
$\varphi_1$ and $\varphi_2$ are the modes that locally diagonalize
the quadratic Lagrangian of perturbations. The blue ellipsoid gives
the typical size of quantum fluctuations.}
\end{center}
\end{figure}

Fortunately we can answer our question in general, with no reference to the actual system that is driving inflation. To this purpose it is particularly convenient to work with the
effective theory for adiabatic scalar perturbations of a generic
FRW Universe. This framework has been developed in
Ref.~\cite{starting}, to which we refer for details. The idea is to
focus on a scalar excitation that is present in virtually all
expanding Universes: the Goldstone boson of broken
time-translations. That is, given the background solution for the
matter fields $\psi_m (t)$, we consider the matter fluctuation
\be \label{def_pi}
\delta \psi_m (x) \equiv \psi_m \big( t + \pi(x) \big) - \psi_m(t) 
\ee
parameterized by $\pi(x)$, and the corresponding scalar perturbation
of the metric as enforced by Einstein equations (after fixing, e.g., Newtonian gauge). This fluctuation
corresponds to a common, local shift in time for all matter fields
and is what in the long wavelength limit is called an `adiabatic'
perturbation. As for all Goldstone bosons, its Lagrangian is largely
dictated by symmetry considerations. This is clearly the relevant
degree of freedom one has to consider to decide whether eternal
inflation is taking place or not. Minimally, a {\em sufficient}
condition for having eternal inflation is to have large quantum
fluctuations back and forth along the classical trajectory. 
In the presence of several matter fields other  fluctuation
modes will be present. For the moment we concentrate on the
Goldstone alone. As we will see at the end of this section, our conclusions are unaltered by the presence of large mixings between $\pi$ and
extra degrees of freedom. The situation is schematically depicted in
Fig.~\ref{pi_figure}.

Of course with dynamical gravity time-translations are gauged and formally there is no Goldstone boson at all---it is ``eaten'' by the gravitational degrees of freedom and one can always fix the gauge $\pi (x)=0$ (`unitary gauge'). Still it remains a convenient parametrization of a particular scalar fluctuation at short distances, shorter than the Hubble scale, which plays the role of the graviton Compton wavelength. This is completely analogous to the case of massive gauge theories, where the dynamics of longitudinal gauge bosons is well described by the ``eaten'' Goldstones at energies higher than the mass.

This approach allows us to analyze essentially any model of inflation.
The reason is that, no matter what the underlying model is, it produces some
$a(t)$, and in unitary gauge the effective Lagrangian breaks time diffs but as we will see is still quite constrained by preserving spatial diffs, so a completely general model can be characterized in a systematic derivative expansion with only a few parameters. The inside-horizon dynamics of the ``clock'' field can be simply obtained from the unitary gauge Lagrangian
by re-introducing the time diff Goldstone {\em \`a la} St\"uckelberg.

The construction of the Lagrangian for $\pi$ is greatly simplified
in `unitary' gauge, $\pi = 0 $. That is, by its very definition eq.~(\ref{def_pi}),
$\pi(x)$ can always be gauged away from the matter sector through a
time redefinition, $t \to t- \pi(x)$. Then the scalar fluctuation
appears only in the metric, thus its Lagrangian only involves the
metric variables. We can reintroduce $\pi$ at any stage of the
computation simply by performing the opposite time diffeomorphism $t
\to t + \pi(x)$. Notice that by construction $\pi$ has dimension of length. All Lagrangian terms must be invariant under the
symmetries left unbroken by the background solution and by the
unitary gauge choice. These are time- and space-dependent spatial
diffeomorphisms, $x^i \to x^i + \xi^i(t,\vec x)$. At the lowest
derivative level the only such invariant is $g^{00}$. Notice that,
given the residual symmetries, the Lagrangian terms will have
explicitly time-dependent coefficients. From the top-down viewpoint
this time-dependence arises because we are expanding around the
time-dependent background matter fields $\psi_m(t)$ and metric $a
(t)$. Because of this, we expect the typical time-variation rate to
be of order $H$, so that at frequencies larger than $H$ it can be
safely ignored.

The matter Lagrangian in unitary gauge takes the form
\cite{starting} \be  \label{S_matter} S_{\rm matter} = \int d^4 x
\sqrt{-g}\left [ \frac{1}{8\pi G} \dot H \,  g^{00} - \frac{1}{8\pi
G} (3 H^2 +\dot H) +  F\big(g^{00}+1 \big) \right]  \ee where
the first two terms are fixed by imposing that the background $a
(t)$ solves Friedmann equations, since they contribute to `tadpole'
terms. $F$ instead can be a generic function that starts quadratic
in its argument $\delta g^{00} \equiv g^{00} + 1$, so that it
doesn't contribute to the background equations of motion, with
time-dependent coefficients, \be \label{F} F(\delta g^{00}) = M^4(t)
\,  (\delta g^{00}) ^2 + \tilde M^4(t) \, (\delta g^{00}) ^3 + \dots
 \ee To match this description with a familiar situation,
consider for instance the case of an ordinary scalar $\phi$ with a
potential $V$ driving the expansion of the Universe. If we perturb
the scalar and the metric around the background solution
$\phi_0(t)$, $a( t)$ and choose unitary gauge, $\phi(x) =
\phi_0(t)$, the Lagrangian  is \be S =  \int d^4 x \sqrt{-g} \Big[
-\sfrac12 \, g^{\mu\nu} \, \di_\mu \phi \di_\nu \phi - V(\phi)
\Big] =  \int d^4 x \sqrt{-g}\Big [ - \sfrac12 \dot \phi_0 ^2 \,
g^{00}- V\big (\phi_0(t) \big) \Big]   \ee which, upon using the
background Friedmann equations, reproduces exactly the first two
terms in eq.~(\ref{S_matter}). Therefore an ordinary scalar
corresponds to the case $F( \delta g^{00}) =0$.

We can now reintroduce the Goldstone $\pi$. This amounts to
performing in eq.~(\ref{S_matter}) the time diffeomorphism \be t \to
t+\pi  \qquad g^{00} \to -1 - 2 \dot \pi + (\di \pi)^2  \ee
Notice that we should really evaluate all explicit functions of time
like $H$, etc., at $t+\pi$ rather than at $t$. However, after
expanding in $\pi$, this would give rise only to non-derivative
terms suppressed by $H$, $\dot H$, etc., that can be safely
neglected as long as we consider frequencies faster than $H$. Of
course in the end we are interested in the physics at freeze-out,
i.e.~exactly at frequencies of order $H$. A correct analysis should
then include these non-derivative terms for $\pi$, as well as the
effect of mixing with gravity---the Goldstone is a convenient
parameterization only at high frequencies. However, being only
interested in orders of magnitude we can use the high-frequency
Lagrangian for $\pi$ and simply extrapolate our estimates down to
frequencies of order $H$.
 From eq.~(\ref{S_matter}) we get
\bea {\cal L}_{\pi} & = & \mpl ^2 \dot H \, (\di \pi)^2 + F \big(- 2
\dot
\pi + (\di \pi)^2 \big )  \\
& = &  (4 M^4 - \mpl ^2 \dot H) \, \dot \pi^2 + \mpl ^2 \dot H \,
(\vec \nabla \pi)^2 + \mbox{higher orders}  \eea where we
neglected a total derivative term and we expanded $F$ as in
eq.~(\ref{F}). At the lowest derivative level, the quadratic
Lagrangian for $\pi$ only has one free parameter, $M^4$. The only
constraint on $M^4$ is that it must be positive for the propagation
speed of $\pi$ fluctuations (the `speed of sound', from now on) $c^2
\equiv \mpl ^2 | \dot H|/(4 M^4 + \mpl ^2 |\dot H|)$ to be smaller
than one. For instance, a relativistic scalar with $c^2=1$
corresponds to $M^4=0$; a perfect fluid with constant equation of
state $0< w <1$ corresponds to $M^4 =  \mpl^2 |\dot H|  \, (1-w)/w$.

If $M^4 \lesssim \mpl^2 |\dot H|$ the speed of sound is of order one
and we can repeat exactly the same analysis as in the case of slow
roll inflation, modulo straightforward changes in the notation.
Therefore, let us concentrate on the case $c^2 \ll 1$, $M^4 \gg
\mpl^2 |\dot H|$; the Lagrangian further simplifies to \be {\cal
L}_{\pi} = 4 M^4  \, \dot \pi^2 + \mpl ^2 \dot H \,  (\vec \nabla
\pi)^2 + \mbox{higher orders}  \label{pi_lagrangian} \ee We now
want to use the Lagrangian (\ref{pi_lagrangian}) to estimate the
size of quantum fluctuations, and to impose that they don't overcome
the classical evolution of the system. For the latter requirement
the $\pi$ language is particularly convenient: $\pi$ is the
perturbation of the classical `clock' $t$, directly in time units,
so we just have to impose $\dot \pi \ll 1$ at freeze-out, that is at
frequencies of order $H$. Alternatively, in unitary gauge we can
look at the dimensionless perturbation in the metric, \be \zeta
\equiv \frac{\delta a}{a} = H \, \pi  \ee so that imposing $\zeta
\ll 1$ at freeze-out we get the same condition for $\pi$ as above.

The typical size of the vacuum quantum fluctuations for a non-relativistic, canonically normalized field $\phi$ with a generic
speed of sound $c \sim \omega/k$ at frequencies of order $\omega$ is
\be \label{canonical_fluct} \langle \phi^2 \rangle _\omega \sim
\frac{k^3}{\omega} \sim \frac {\omega^2}{c^3}  \ee where the
$\omega$ in the denominator comes from the canonical wave-function
normalization, and the $k^3$ in the numerator from the measure in
Fourier space. Taking into account the non-canonical normalization
of $\pi$, at frequencies of order $H$ we have \be \label{pi_fluct}
\langle \pi^2 \rangle_H \sim \frac{H^2}{M^4 \, c^3}   \ee The
size of quantum fluctuation is enhanced for smaller sound speeds
$c$. And since $c^2$ is proportional to $|\dot H|$, clearly there
will be a lower bound on $|\dot H|$ below which the system is
eternally inflating. Indeed imposing $\langle \dot \pi ^2 \rangle_H
\ll 1$  and using $c^2 = \mpl ^2 |\dot H| / M^4$ we directly get 
\be \label{1overc}
|\dot H| \gg \frac{1}{c} \, G H^4  
\ee 
which in the limit $c \ll 1$ is even stronger than eq.~(\ref{Hdot}).
From this the constraint $dS \gg \frac1c \, dN$ immediately follows.

This proves our bounds for all models in which the physics of
fluctuations is correctly described by the Goldstone two-derivative
Lagrangian, eq.~(\ref{pi_lagrangian}). This class includes for
instance all single-field inflationary models where the Lagrangian
is a generic function of the field and its first derivatives, ${\cal
L} = P\big((\di \phi)^2, \phi\big)$, from slow-roll inflation to k-inflation models \cite{k}. It is however useful to consider an even
stronger bound that comes from taking into account non-linear interactions
of $\pi$. This bound will be easily generalizable to theories with
sizable higher-derivative corrections to the quadratic $\pi$
Lagrangian, like the ghost condensate. This is where the null energy
condition comes in.

The null energy condition requires that the stress-energy tensor
contracted with any null vector $n^\mu$ be non-negative, $T_{\mu\nu}
\, n^\mu n^\nu \ge 0$. We can read off the stress energy tensor from
the matter action in unitary gauge eq.~(\ref{S_matter}) by
performing the appropriate derivatives with respect to the metric.
Given a generic null vector $n^\mu = (n^0, \vec n)$ the relevant
contraction is \be \label{NEC} T_{\mu\nu} \, n^\mu n^\nu = -2 \,
(n^0)^2 \big[ \mpl^2 \dot H + F' ( \delta g^{00} ) \big]  \ee
where $\delta g^{00} = g^{00}+1$ is the fluctuation in $g^{00}$
around the background. In a more familiar notation, for a scalar
field with a generic Lagrangian ${\cal L} = P(X, \phi)$, $X \equiv
(\di \phi)^2$, the above contraction is just $T_{\mu\nu} \, n^\mu
n^\nu =  2 \, (n^\mu \, \di_\mu \phi) ^2 \, \di_X P$, so the NEC is
equivalent to $\di_X P \ge 0$.

\begin{figure}[t!]
\begin{center}
\includegraphics[width=10cm]{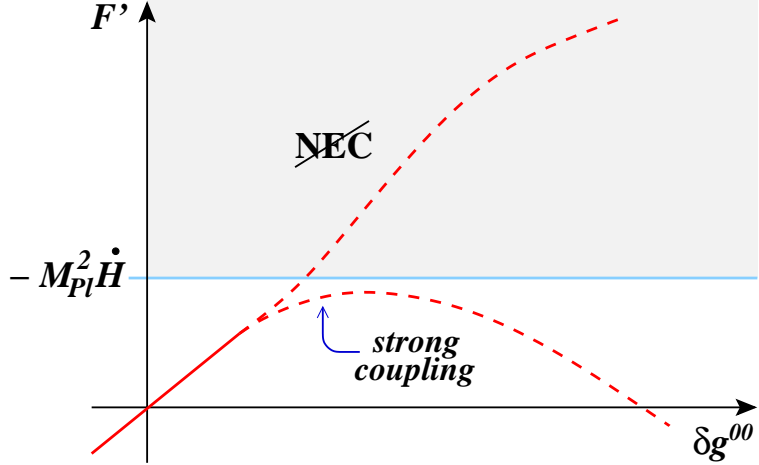}
\caption{\small \it \label{F_fig} The null energy condition is violated
whenever $F'(\delta g^{00})$ enters the shaded region, $F' +
M^2_{Pl} \dot H
> 0$.
Since $F'$ starts with a strictly positive slope at the origin, to
avoid this one needs that higher derivatives of $F$ bend $F'$ away
from the NEC-violating region. The smaller $|\dot H|$, the stronger
the needed `bending'. This can make $\pi$ fluctuations strongly
coupled at $H$. }
\end{center}
\end{figure}

On the background solution $\delta g^{00} $ vanishes and since
$F'(0) $ vanishes by construction, the NEC is satisfied---of course
as long as $\dot H$ is negative, as we are assuming. However $F''(0)
= M^4$ is positive, making $F'$ positive for positive $\delta g^{00}
$. As a consequence the r.h.s.~of eq.~(\ref{NEC}) is pushed towards
negative values for positive $\delta g^{00} $. So the NEC tends to
be violated in the vicinity of the background solution unless higher
order terms in the expansion of $F$, eq.~(\ref{F}), save the day,
see Fig.~\ref {F_fig}. But this can only happen if their coefficient
is large enough. For instance in order for the $n$-th order term to
keep eq.~ (\ref{NEC}) positive definite its coefficient must be at
least as large as $M^4 \, (M^4 / \mpl^2 \dot H)^{n-2}$. The smaller
$|\dot H| $, the closer is  the background solution to violate the
NEC, and so the larger is the `correction' needed not to violate it.
But then if higher derivatives of $F$ on the background solution are
large, self-interactions of $\pi$ are strong. Minimally, we don't
want $\pi$ fluctuations to be strongly coupled at frequencies of
order $H$. If this happened the semiclassical approximation would
break down, and the classical background solution could not be
trusted at all---quantum effects  would be as important as the
classical dynamics in determining the evolution of the system, much
like in the usual heuristic picture of eternal inflation.

Recall that the argument of $F$ expressed in terms of the Goldstone
is $\delta g^{00} = -2 \dot \pi -\dot \pi^2 +(\vec \nabla \pi)^2$.
Given an interaction term $(\delta g^{00})^n$, it is easy to check
that for fixed $n$ the most relevant $\pi$ interactions come from
taking only the linear $\dot \pi$ term in $\delta g^{00}$, i.e.~$
(\delta g^{00})^n \to \dot \pi ^n$. Therefore, if eq.~(\ref{NEC}) is
kept positive definite thanks to the $n$-th order term in the Taylor
expansion of $F$, the ratio of the $\pi$ self-interaction induced by
this term and the free kinetic energy of $\pi$ is \be \label{ratio}
g_n \equiv \frac{M^4 \, (\, M^4 / \mpl^2 |\dot H| \, )^{n-2} \,
\dot \pi^n}{M^4 \, \dot \pi^2} = \left( \frac{M^4 \, \dot
\pi}{\mpl^2 | \dot H|} \right)^{n-2} \sim  \left( \frac{M^2 \,
H^2}{\mpl^2 |\dot H| \cdot c^{3/2}} \right)^{n-2} = \left(
\frac{H^4}{\mpl^2 |\dot H| \cdot c^5} \right)^{\frac{n-2}{2}} 
\ee where  we plugged in the size of typical quantum fluctuations at
frequencies of order $H$, eq.~(\ref{pi_fluct}), and we used the fact
that $\mpl^2 |\dot H| = c^2 \, M^4$.
 From eq.~(\ref{ratio}) it is evident that if we require that quantum
fluctuations be weakly coupled at frequencies of order $H$ we
automatically get the constraint \be \label{final_bound} |\dot H|
\gg \frac{1}{c^5} \, G H^4 \; , \qquad dS \gg \frac{1}{c^5} \, dN
\ee on the background classical solution.

The above proof holds in all cases where the Goldstone
two-derivative Lagrangian, eq.~(\ref{pi_lagrangian}) is a good
description of the physics of fluctuations. However, when $| \dot
H|$ is very small the $ (\vec \nabla \pi)^2$ term appears in the
Lagrangian with a very small coefficient, and one can worry that
higher derivative corrections to the $\pi$ quadratic Lagrangian
start dominating the gradient energy. This is exactly what happens
in ghost inflation, where the $(\vec \nabla \pi)^2$ term is
absent---in agreement with the vanishing of $ \dot H$---and the
spatial-gradient part of the quadratic Lagrangian is dominated by
the $(\nabla^2 \pi)^2$ term, which enters the Lagrangian with an
arbitrary coefficient  \cite{ghost_C,ghost_I}. In such cases, at all
scales where the gradient energy is dominated by higher derivative
terms one has $\mpl^2 |\dot H| < c^2 \, M^4$, where $c$ is the
propagation speed, simply because the $(\nabla \pi)^2$ term of 
eq.~(\ref{pi_lagrangian}) is not the dominant source of gradient energy,
thus the sound speed is dominated by other sources. So the last
equality in eq.~(\ref{ratio}) becomes a `$>$' sign, and our bound
gets even stronger. Therefore our results equally apply to theories
where higher derivative corrections can play a significant role,
like the ghost condensate.

In summary: imposing that the NEC is not violated in the vicinity of
the background solution implies sizable non-linearities in the
system. For smaller $|\dot H|$ the system is closer to violating the
NEC---$\dot H = 0$ saturates the NEC. So the smaller $|\dot H|$, the
larger the non-linearities needed to make the system healthy.
Requiring that fluctuations not be strongly coupled at the scale $H
$---a necessary condition for the applicability of the semiclassical
description---sets a lower bound on $|\dot H|$,
eq.~(\ref{final_bound}).

So far we neglected possible mixings of $\pi$ with other light fluctuation modes.
However our conclusion are unaltered by the presence of such mixings. At any give moment of time $t$ the quadratic Lagrangian for fluctuations can be diagonalized,
\be \label{lagr_phi} 
{\cal L} = \sfrac12 \sum_{i=1}^N \dot \varphi_i ^2 - c_i^2 (\vec \nabla \varphi_i)^2 
\ee
Typical quantum fluctuations now define an ellipsoid in the $\varphi_i$'s space, whose semi-axes depend on the individual speeds $c_i$ (see Fig.~\ref{pi_figure}). The Goldstone $\pi$ corresponds to some specific direction in field space, and in any direction quantum fluctuations are bounded from below by the shortest semi-axis.
By requiring that the system does not enter eternal inflation it is straightforward to show that our bound (\ref{1overc}) generalizes to 
\be 
|\dot H| \gg \frac{1}{c_{\rm max}} \,
G H^4  \; , \qquad dS \gg \frac {1}{c_{\rm max}} \, dN  
\ee
where $c_{\rm max}\le 1$ is the maximum of the $c_i$'s.
The generalization to theories in which higher spatial derivative
terms are important proceeds along the same lines as in the case of
the $\pi$ alone, by imposing that the NEC is not violated along
$\pi$ and that $\pi$ fluctuations are not strongly coupled at $H$.



\section{Null energy condition and thermodynamics of horizons}
\label{sec:NEC}

The proof of our central result (\ref{dS}) and the related interpretation of what finite 
de~Sitter entropy means crucially relies on the null energy condition,
\be
\label{nec}
T_{\mu\nu}n^\mu n^\nu\geq 0
\ee
where $n^\mu$ is null. The history of general relativity knows many examples 
when the assumed ``energy conditions"---assumptions about the properties
of physically allowed energy-momentum tensors---turned out to be wrong.
In the end, the NEC  is also known to be violated both by quantum effects (Casimir energy, Hawking
evaporation) and by non-perturbative objects (orientifold planes in string theory).
So it is important to clarify to what extent the violation of the NEC needed to get around the bound 
(\ref{dS}) is qualitatively different from these examples, and why the relevance of the NEC in our proof is  more than just a technicality.

Note first that all qualitative arguments of section~\ref{sec:locality}, indicating that sharply defined local observables are absent in quantum gravity, implicitly rely  on the notion of positive gravitational energy. Indeed, schematically these arguments reduce to saying that,
by the uncertainty principle, preparing arbitrarily precise clocks and rods requires concentrating
indefinitely large energy in a small volume. Then the self-gravity of  clocks and rods themselves causes the volume
to collapse into a black hole and screws up the result of the measurement. Clearly this problem
would not be there if there were some negative gravitational energy available around.
Using this energy one would be able to screen the self-gravity of clocks and rods and to perform
an arbitrarily precise local measurement. 
NEC is a natural candidate to define what the positivity of energy means; at the end  it is the only energy condition in gravity that
cannot be violated by just changing the vacuum part of the energy-momentum, $T_{\mu\nu}\to T_{\mu\nu}+\Lambda g_{\mu\nu}$. 
Indeed the NEC is a crucial assumption in proving the positivity of the ADM  mass in asymptotically flat spaces \cite{SchonYau,WittenMass}.

Generically, classical field theoretic systems violating NEC suffer from either ghost or rapid gradient instabilities. 
In a very broad class of systems, including conventional relativistic fluids, these instability can be proven \cite{fluids} to originate
from the``clock and rod" sector of the system---one of the Goldstones of the spontaneously broken space-time translations is either a ghost or
has an imaginary propagation speed. 
For instance, if space translations are not spontaneously broken and only the Goldstone of time translations  (the ``clock" field $\pi$
of section \ref{sec:proof}) is present, then the instability is due to the wrong-sign gradient energy
in the Goldstone Lagrangian~(\ref{pi_lagrangian}) in the NEC violating case $\dot{H}>0$. The examples of stable NEC violations we mentioned above
avoid this problem by either being quantum and non-local effects
(Casimir energy and Hawking process) or by projecting out the corresponding Goldstone mode
(orientifold planes). This allows to avoid the instability, but simultaneously makes these systems incapable of providing the non-gravitating clocks
and rods.

Nevertheless, stable effective field theories describing non-gravitating systems of clocks and rods can be constructed. 
This is the ghost condensate model \cite{ghost_C} where space diffs are unbroken, and so only the clock field appears, as
well as more general models describing gravity in the Higgs phase where Goldstones of the space diffs are present as well \cite{Rubakov:2004eb,Dubovsky:2004sg}.
All these setups provide constructions of de~Sitter space with intrinsic clock variable and thus allow to get around our bound (\ref{dS}).
Related to that, all these theories describe systems on the verge of violating NEC, and small perturbations around their vacuum violate it.
Nevertheless these effective theories avoid rapid instabilities as a combined result of
taking into account  the higher derivative operators in the Goldstone sector and of imposing  special symmetries.

Does the existence  of these counterexamples cause problems in relating the bound (\ref{dS}) to the fundamental properties of de~Sitter space in quantum gravity? We believe that the answer is no, and that actually the opposite is true---this failure of the bound (\ref{dS}) 
provides a quite non-trivial support to the idea that the bound is deeply related to de~Sitter thermodynamics. The reason is that the conventional black hole thermodynamics also fails in these models \cite{Dubovsky:2006vk}. 

To see how this can be possible, note that, more or less by construction, all these models spontaneously break Lorentz invariance. For instance, in the ghost condensate Minkowski or de~Sitter vacuum a non-vanishing 
time-like vector---the gradient of the ghost condensate field $\d_\mu\phi$---is present.  As usual in Lorentz violating theories the maximum 
propagation velocities need not be universal for different fields, now as a consequence of the direct interactions with
the ghost condensate. Being a consistent non-linear effective theory, ghost condensate allows to study the consequences of the velocity differences
in a black hole background. The result is very simple---the effective metric describing propagation of a
field with $v\neq 1$ in a Schwarzschild background has the Schwarzschild form with a different value of the mass.
As one could have expected, the black hole horizon appears larger for subluminal particles and smaller for superluminal ones.
As a consequence, the temperature of the Hawking radiation is not universal any longer; ``slow" fields are radiated with lower
temperatures than ``fast" fields.

\begin{figure}[t!]
\begin{center}
\includegraphics[width=6cm]{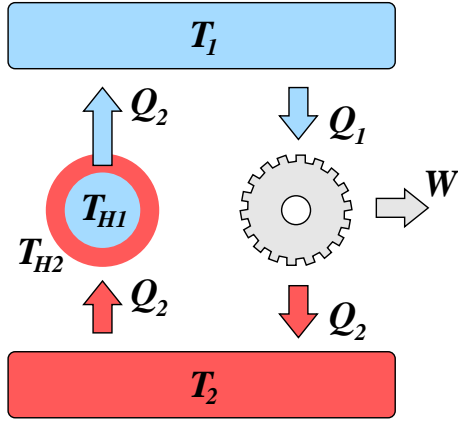}
\caption{\label{Carnot} \small \it In the presence of the ghost condensate black holes can have different temperatures for different fields. This allows to perform thermodynamic transformations whose net effect is the transfer of heat $Q_2$ from a cold reservoir at temperature $T_2$ to a hotter one at temperature $T_1$ {\em (left)}. 
Then one can close a cycle by feeding heat $Q_1$ at the higher temperature
$T_1$ into a machine that produces work $W$ and as a byproduct releases
heat $Q_2$ at the lower temperature $T_2$ {\em (right)}. 
The net effect of the cycle is the conversion of heat into mechanical work.}
\end{center}
\end{figure}

Also the horizon area does not have a universal meaning any longer, making it impossible to define the black hole
 entropy just as a function of mass, angular momentum and gauge charges. To make the conflict with thermodynamics explicit, let us consider a black hole radiating two different
non-interacting species with different Hawking  temperatures  $T_{H1}>T_{H2}$. Let us bring the black hole in thermal contact
with two thermal reservoirs containing species 1 and 2 and having temperatures $T_1$ and $T_2$ respectively.
By tuning these temperatures one can arrange that they satisfy
\[
T_{H1}>T_1>T_2>T_{H2}
\]
and the thermal flux from the black hole to the first reservoir is exactly equal in magnitude to the flux from the second reservoir to the black hole.
As a result the mass of the black hole remains unchanged and the heat is transferred from the cold to the hot body in contradiction with the second
law of thermodynamics, see Fig.~\ref{Carnot}.

The case for violation of the second law of black hole thermodynamics in models with spontaneous Lorentz violation is even strengthened by the observation
\cite{Eling:2007qd} that the same conclusion can be achieved purely at the classical level and without neglecting the interaction between the two species.
This classical process is analogous to the Penrose process. Namely, in a region between the two horizons  the energy of the ``slow" field
can be negative similarly to what happens  in the ergosphere of a Kerr
black hole. The fast field can escape from this region making it possible to arrange an analogue of the Penrose process. In the case at hand, this process 
just extracts energy from the black hole by decreasing its mass. The mass decrease can be compensated by throwing in more entropic stuff, which again 
results in an entropy decrease outside with the black hole parameters remaining unchanged (this does not happen in the conventional  Penrose process
because the angular momentum of the black hole changes).

Actually, it is not surprising at all that a violation of the NEC implies the breakdown of black hole thermodynamics, as the NEC
is needed in the proof \cite{Flanagan:1999jp} of  the covariant entropy bound \cite{Bousso:1999xy}, which is one of the basic ingredients of black hole thermodynamics and holography.
Also note that the above  conflict with thermodynamics is just a consequence of spontaneous breaking of Lorentz invariance (existence of non-gravitating clocks);
in particular, it is there even if one assumes that all fields propagate subluminally. 

The second law of thermodynamics is a consequence of a few
very basic properties, such as unitarity, so it is expected to hold in any sensible quantum theory. Hence, the only chance for Lorentz violating
models to be embedded in a consistent microscopic theory is if black holes are not actually black in these theories, so that the observer can
measure both the inside and the outside entropy and there is no need for a purely outside counting as provided by the Bekenstein formula (this is indeed what happens if space diffs are broken as well, due to the existence of instantaneous interactions). In any case, this definitely puts the
ghost condensate with other Lorentz violating models
in a completely different ballpark  from GR as far as the physics of horizons goes. That is why we find it encouraging for a thermodynamical interpretation
of the bound (\ref{dS}) that is also violated by the ghost condensate.



\section{Open questions}

We have seen that all NEC-obeying models of inflation that do not eternally inflate increase the de~Sitter at a minimal rate, $dS/dN \gg 1$, and therefore cannot sustain an approximate de~Sitter phase for longer that $N \sim S$ $e$-foldings. This gives an observational way of determining whether or not inflation is eternal. For instance, if our current accelerating epoch lasts for longer than $\sim 10^{130}$ years, or if $(1 + w)$ is smaller than $10^{-120}$, our current inflationary epoch is eternal. While these are somewhat challenging measurements, they can at least be done at timescales shorter than the recurrence time! 

This bound implies that an observer exiting into flat space in the asymptotic future cannot detect more than $e^S$ independent modes coming from inflation, which
matches nicely with the idea of de~Sitter space having a finite-dimensional Hilbert space of dimension $\sim e^S$.
Although we are not able to provide a microscopic counting of de~Sitter entropy, we can at least give an {\em operational} meaning to the number of de~Sitter degrees of freedom. The NEC is very important in proving our bound; indeed
the NEC is crucial in existing derivations of various holographic bounds, and indeed consistent EFTs that violate the NEC like the ghost 
condensate are also known to violate the thermodynamics of black hole horizons.  This suggests that our bound is related to holography.

\begin{figure}[t!]
\begin{center}
\includegraphics[width=13cm]{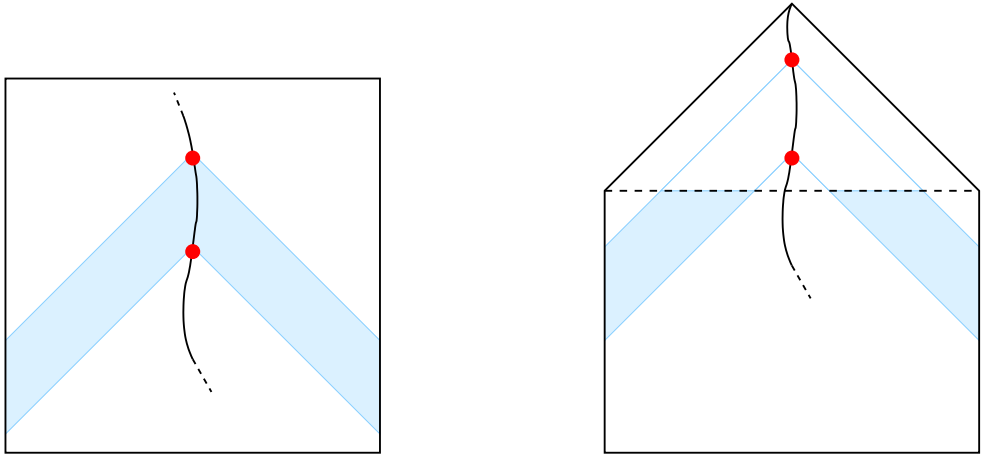}
\caption{\label{covariant}\small \it A possible covariant generalization of our bound. Given an observer's worldline and a``start'' and an ``end'' times {\em (red dots)}, one identifies the portion of de Sitter spacetime that is detectable by the observer in this time interval {\em (shaded regions)}.
Then EFT properly describes such a region only for spacetime volumes smaller than $\sim e^S H^{-4}$.
If applied to eternal de Sitter {\em (left)} this gives the Poincar\'e recurrence time $e^S H^{-1}$ times
the causal patch volume $H^{-3}$.
If applied to an FRW observer after inflation {\em (right)} it gives $S$ $e$-foldings times $e^{S}$ Hubble volumes.
}
\end{center}
\end{figure}

We can view different inflationary models as possible regularizations of pure de~Sitter space in which a semiclassical analysis in terms of a local EFT is reliable. Then our universal bound suggests that {\em any} semiclassical, local description of de~Sitter space cannot be trusted past $\sim S$ Hubble times and further than $\sim e^S$ Hubble radii in space---perhaps a more covariant statement is that the largest four-volume one can consistently describe in terms of a local EFT is of order $e^S H^{-4}$, see Fig.~\ref{covariant}.
Notice that this is analogous to what happens for a black hole: in order not to violate unitarity the EFT description {\em must} break down after a time of order $S$ Schwarzschild times, when more than $e^S$ modes must be invoked behind the horizon to accommodate the entanglement entropy.

Ultimately we are interested in eternal inflation, in particular in its effectiveness in populating the string landscape.
In this case the relevant mechanism is false vacuum eternal inflation, in which there is no classically rolling scalar to begin
with, and the evolution of the Universe is governed by quantum tunneling.  Our analysis does not directly apply here---there is no classical non-eternal version of this kind of inflation. In particular, in the slow roll eternal inflation case an asymptotic
future observer only has access to the late phase of inflation, when the Universe is {\em not} eternally inflating. The eternal inflation part corresponds to density perturbations of order unity, thus making the Hubble volume surrounding the observer collapse when they become observable. As a consequence the number of possible independent measurements such an observer can make is always bounded by $e^S$. 

In the false vacuum eternal inflation case instead there can be asymptotic observers who live in a zero cosmological constant bubble. This is the case if the theory does not have negative energy vacua, or if the zero energy ones are supersymmetric, and therefore perfectly stable. Such zero-energy bubbles are occasionally hit from outside by small bubbles that form in their vicinity, but these collisions are not very energetic and do 
not perturb significantly the bubble evolution---the total probability of being hit and eaten by a large bubble is small, of order $\Gamma H^{-4} \ll 1$, where $\Gamma$ is the typical transition rate per unit volume. By measuring the remnants of such collisions the observer inside the bubble can gather information about the outside de~Sitter space and the landscape of vacua \cite{MaShSu}. Then, in this case 
these measurements play the same role in giving an operational definition of de~Sitter degrees of freedom
as density perturbations did in slow-roll inflation.  
\begin{figure}[t!]
\begin{center}
\includegraphics[width=13cm]{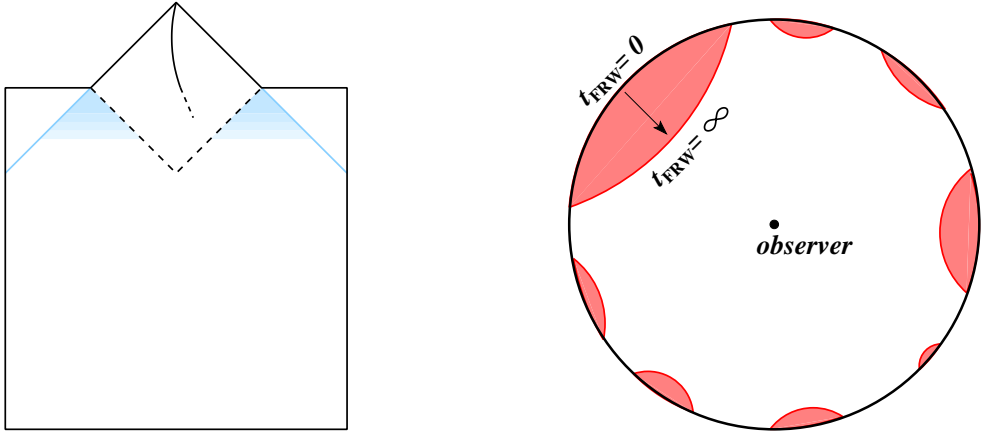}
\caption{\label{bubble} \small \it {\em (Left)} In false vacuum eternal inflation there seems to be no limit to the spacetime volume of the outside de Sitter space an asymptotic flat-space observer can detect. The spacetime volumes diverges in the shaded corners. Bubble collisions don't alter this conclusion; the pattern of collisions is simply depicted on a Poincar\'e disk representation  of the hyperbolic FRW spatial slices {\em (Right)}. Maloney, Shenker and Susskind argue that observers in the bubbles can make an infinite number of observations and arrive at sharply defined observables.
}
\end{center}
\end{figure}
But now there seems to be no limit to how many independent measurements an asymptotic observer can make. The expected total number of bubble collisions experienced by a zero-energy bubble is infinite, and with very good probability none of these collisions destroys the bubble. It is true that as time goes on for such an observer it becomes more and more difficult to perform these measurements---collisions get rarer and rarer, and their observational consequences get more and more redshifted. Still we have not been able to find a physical reason why these observations cannot be done, at least in principle. The asymptotic observer in the bubble can in principle perform infinitely many independent measurements, and Maloney, Shenker and Susskind argue that these might give sharply defined observables \cite{MaShSu}. The case of collisions with negative vacuum energy supersymmetric bubbles is particularly interesting; in this case, as the boundary of the zero energy bubble is covered by an infinite fractal of domain-wall horizons \cite{FHS}, the pattern of bubble collisions with other supersymmetric vacua as seen on the hyperbolic spatial slices of the bubble FRW Universe is shown in Fig.~(\ref{bubble}) where the hyperbolic space is represented as a Poincar\'e disk; at early times the walls are at the boundary while at infinite time they asymptote to fixed Poincar\'e co-ordinates as shown. The pattern of collisions is scale-invariant, reflecting the origin of the bubbles in the underlying de~Sitter space. Still, it appears that an observer away from these walls can make an infinite number of observations. This apparently violates the expectation that one should not be able to assign more than $e^S$
independent states to de~Sitter space. Perhaps false vacuum eternal inflation is a qualitatively different regularization of de~Sitter space than offered by the class of inflationary models we studied for our bound. There may be some more subtle effect that prevents the bubble observer from making observations with better than $e^{-S}$ accuracy of the ambient de~Sitter space. Or perhaps the limitation is correct, and it is the effective field theory description that is breaking down when more than $e^S$ observations are allowed, much as in black hole evaporation. We believe these issues deserve further investigation.

\section*{Acknowledgments}
We thank Tom Banks, Raphael Bousso, Ben Freivogel, Steve Giddings, David Gross, Don Marolf, Joe Polchinski, Leonardo Senatore, and Andy Strominger for stimulating discussions. We especially thank Juan Maldacena for clarifying many aspects of the information paradox and de Sitter entropy, and also Alex Maloney and Steve Shenker for extensive discussions of their work in progress with Susskind.  The work of Enrico Trincherini is supported by an INFN postdoctoral fellowship.


\end{document}